\documentstyle[ltwol,psfig]{article}

\def\be{\begin{equation}}
\def\ee{\end{equation}}
\def\bea{\begin{eqnarray}}
\def\eea{\end{eqnarray}}

\begin{document}

\title{MAJORANA NEUTRINO TRANSITION MAGNETIC MOMENTS \\
IN LEFT-RIGHT SYMMETRIC MODELS}

\author{M. CZAKON, J. GLUZA  and M. ZRA\L EK}

\address{
Department of Field Theory and Particle Physics \\
Institute of Physics, University of Silesia, Uniwersytecka 4\\
PL-40-007 Katowice, Poland \\
E-mails:czakon,gluza,zralek@us.edu.pl}


\twocolumn[\maketitle\abstracts{
Transition magnetic moments of Majorana neutrinos are discussed in the 
frame of the most natural version of the
LR model (with left- and right-handed triplets and a bidoublet in the Higgs
sector). We show that their largest values could be 
at most $6\cdot 10^{-13} \mu_B$ 
from diagrams with $W_L$  in the loop. This could happen for specific models
where (i)
neutrino-charged lepton mixing is maximal and (ii) $\kappa_1 \simeq
\kappa_2$ (VEVs for neutral Higgs fields in the bidoublet $\phi$ are equal).
Contributions from diagrams with charged Higgses 
in the loop are smaller than those in the SM with right-handed neutrinos. }]

\section{Introduction}

\subsection{Bounds on $\nu$-magnetic moments} 
The existence of a nonzero neutrino magnetic moment is a theoretically 
interesting issue in neutrino physics which is even strengthened by the first
indication that neutrinos are massive particles \cite{sup}. 
Whether it is also an experimentally relevant quantity depends obviously on
its magnitude. 

Measurements of the  ${\nu_i} e^- \rightarrow {\nu_i}
e^-$ 
(${\nu_i}=\bar{\nu}_e,\nu_e,\bar{\nu}_\mu$) and the $\nu_\tau
e^- \rightarrow \nu_\tau e^-$ cross sections give the following limits 
\footnote{These limits are quoted by PDG. The authors know of no
terrestrial experimental limits on transition magnetic moments (Majorana neutrinos),
although there is a general consensus that they are at most of the same
order.}\cite{exp}
\begin{eqnarray}
\mu_{\nu_e} & \leq & 1.8 \cdot 10^{-10} \mu_B, \\
\mu_{\nu_{\mu}} & \leq & 7.4 \cdot 10^{-10} \mu_B, \\
\mu_{\nu_{\tau}} & \leq & 5.4 \cdot 10^{-7} \mu_B.
\end{eqnarray}

There are also astrophysical bounds in addition to  direct laboratory limits
given above.
In particular red giant luminosity and helium stars cooling
by neutrinos emission impose $\nu$-magnetic moments
smaller than $10^{-12} \mu_B$ \cite{astro}.
However, this limit is not as reliable as the terrestrial \cite{pal}.
We expect that values in the range $\mu \sim 10^{-10} \div 10^{-12} \mu_B$
could have practical implications for the Sun \cite{akhm}, Supernova and/or
neutron star physics \cite{fuj}.

Unfortunately, the Standard Model (SM) with its massless 
neutrinos and 
sole left-handed currents leaves no space for a nonvanishing neutrino
magnetic moment.
It is easily understandable. 
Since it arises from the operator $\sigma_{\mu \nu}q^{\nu}$ and as
\begin{equation}
\bar{\Psi}_f\sigma_{\mu \nu}\Psi_i=\bar{\Psi}_{fL}\sigma_{\mu \nu}\Psi_{iR}+
\bar{\Psi}_{fR}\sigma_{\mu \nu}\Psi_{iL},
\end{equation}
we can see that there is a chirality change which makes it necessary to
have both left and right handed particle states.

The easiest way to avoid this shortcoming of the SM is to add right-handed
singlets with additional mass terms. The latter also change chirality.
 
As it was found such a theory yields \cite{mod}
\begin{eqnarray}
\mu_{\nu_\alpha
\nu_\beta}&=&\frac{3eG_F}{16\sqrt{2}\pi^2}(m_{\alpha}+m_\beta)
\sum\limits_{l=e,\mu,\tau}Im \left( U_{\beta l}^\dagger U_{l \alpha} \right)
\left(\frac{m_l}{M_W} \right)^2
 \nonumber \\
& \simeq & 1.6 \times 10^{-19} \left( \frac{m_{\nu_\alpha}+m_{\nu_\beta}}
{1 eV} \right) \nonumber \\
&\times& \sum\limits_{l=e,\mu,\tau}Im \left( U_{\beta l}^\dagger 
U_{l \alpha} \right) \left(\frac{m_l}{M_W} \right)^2.
\end{eqnarray}

Recent Superkamiokande as well as solar neutrino results and
cosmological arguments suggest that neutrino masses are much smaller than 
present terrestrial bounds. 
Since obviously $$\sum\limits_{l=e,\mu,\tau}Im \left( U_{\beta l}^\dagger 
U_{l \alpha} \right) \left(\frac{m_l}{M_W} \right)^2 < 10^{-4}$$ 
we can safely assume that in the very best of the cases 
$\mu_{\nu_{e(\mu)}\nu_\tau} \leq 10^{-16} 
\mu_B$ for $m_{\nu_\tau}$ of order of a few MeV.

This discouraging conclusion shows that we need 
more sophisticated  models than the SM alone 
(with right-handed neutrinos) to get experimentally viable 
magnetic moments.


\subsection{Models with large neutrino magnetic moments}

Plenty of models with large  $\nu$-magnetic
moments have emerged in literature during past decades \cite{pal,mod}. 
Those that remain
interesting from the phenomenological point of view 
can be subdivided into two categories with reference to our
problem (i) renormalizable ones (charged scalars in the SM, left-right symmetric
models,...) and (ii) finite ones (MSSM, supersymmetric left-right model,...).
In the latter class there is a direct connection
between the mass of the neutrino and its magnetic moment which requires
special treatment (see the case of MSSM in \cite{moh1}). The situation turns
up to be much simpler in the former class where corrections to the mass are divergent. Then
renormalization makes it a free parameter. As the magnetic moment contribution
is always finite we can safely consider it alone.
    

So, how to make the magnetic moment contribution  larger
than in the SM?

First we need to generate a term containing no external neutrino mass. To
one loop order this can only be done by left-right transition in vertices
which is possible by adding
\begin{enumerate}
\item charged Higgs scalars, whose Yukawa couplings contain left-right
transition from neutrinos to charged leptons,
\end{enumerate}
and/or
\begin{enumerate}
\item[2.] right currents
\end{enumerate}

Both of these are present in the left-right symmetric models, on which we
shall concentrate in the present article..
Although many estimations have been done by other authors, too rough
treatments led to quite contrary conclusions \cite{large,opp}.
Here we attempt to clarify the situation by extending the calculation to the
Higgs sector and using more phenomenological arguments.

\section{$\nu$-transition magnetic moments in Left-Right symmetric models}

We concentrate on the popular version of the L-R symmetric model with Higgs
bidoublet $\phi$ and two Higgs triplets $\Delta_{L,R}$. In such a model 
neutrinos have a Majorana character meaning only transition magnetic
moments are allowed. Other Higgs sectors may lead to Dirac neutrinos,
however the magnetic moment is very small \cite{nasza}.

Until now only diagrams with gauge bosons in the loop (see
Fig.1) have been considered in the literature, with the dominant 
$W_1$ gauge boson contribution. 

\begin{figure}
\center
\psfig{figure=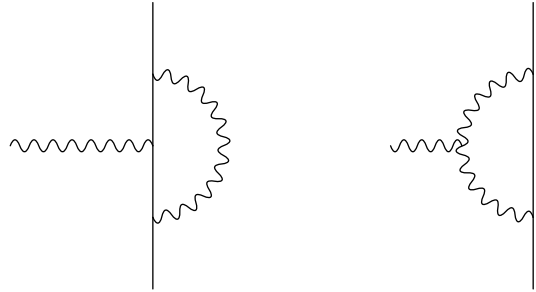,height=1.5in}
\caption{Diagrams for the neutrino magnetic moment with $W_{1,2}$ gauge
bosons and charged leptons (wavy and solid lines, respectively) in the loop. 
External wavy(solid) line(s) stands for the
photon(neutrinos).}
\end{figure}

However, as the charged Higgs - leptons ($H^{\pm}l\nu$)
coupling is proportional to heavy neutrino masses, even if $H^{\pm}$  masses 
are very large the contribution of diagrams with exchanged Higgs particles
to $\mu_\nu$ seems to be interesting.

The contribution to the $\mu_{\alpha \beta}$ can be described by the
following diagrams\footnote{The calculation has been performed 
in the unitary gauge.} (1) with gauge bosons $W^{\pm}_{1,2}$ exchange
(Fig.1), (2)
with charged Higgs bosons $H_{1,2}$ exchange (Fig.2) , (3) with both 
$W^{\pm}_{1,2}$ and $H_{1,2}$ exchange (Fig.3). 

\begin{figure}
\center
\psfig{figure=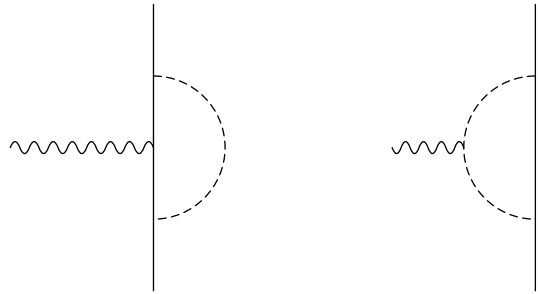,height=1.5in}
\caption{Diagrams for the neutrino magnetic moment with $H_{1,2}$ scalars
(dashed lines)
in the loop. External wavy(solid) line(s) stands for the
photon(neutrinos).}
\end{figure}

\begin{figure}
\center
\psfig{figure=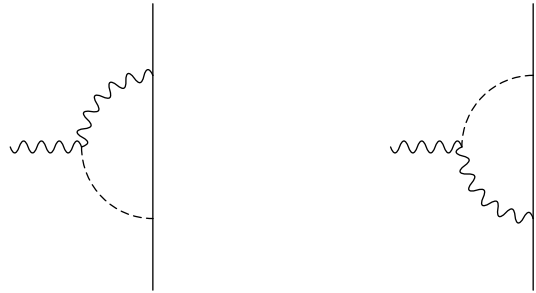,height=1.5in}
\caption{Diagrams for the neutrino magnetic moment with $H_{1,2}$ (dashed
lines) and
$W_{1,2}$ (wavy lines) charged particles 
in the loop. External wavy(solid) line(s) stands for the
photon(neutrinos).}
\end{figure}

\subsection{Diagrams with gauge bosons in the loop}

As $M_{W_2}>>M_{W_1}$
the diagrams with $W_1$ exchange dominate. Their contribution to the 
$\mu_{\alpha \beta}$ can be classified into two categories
\begin{itemize}
\item[(1)] diagrams with neutrino mass insertion on the external neutrino legs,

and

\item[(2)] a diagram with charged lepton mass insertion on the internal
lepton leg.
\end{itemize}
Diagrams from class (1) are proportional to the sum of neutrino masses 
$m_\alpha+m_\beta$ and are the same 
as in Eq.(5). The diagram (2) is proportional to the $W_L-W_R$ mixing 
angle $\xi$, namely \cite{mod}

\begin{eqnarray*}
\mu_{\nu_a \nu_b} & \simeq & \frac{\sqrt{2}G_F}{\pi^2}\sin{\xi}\cos{\xi}
\cdot m_e  \nonumber \\
&& \sum\limits_{\alpha=e,\mu,\tau} m_{\alpha} 
 Im \bigg[ {(K_R)}_{a \alpha }{(K_L^{\dagger})}_{\alpha b} +
{(K_L)}_{a\alpha}{(K_R^{\dagger})}_{\alpha b}\bigg] \mu_B. 
\end{eqnarray*}
\begin{equation}
\end{equation}

The numerical value of $\mu_{\alpha \beta}$ depends on the mixing angle
$\xi$ and the $K_{R(L)}$ matrix elements. If we assume that the $K_L$
matrix is diagonal (lepton number conservation), the $K_R$ matrix elements
are given by the see-saw mechanism
\begin{equation}
K_R \sim O \left( \frac{<m_D>}{M_N} \right)
\end{equation}
and $K_R \sim 0.01$ for $<m_D> \sim 1$ GeV and $m_N=100$ GeV.
Taking also that (limits from fits to low energy data \cite{jas})
\begin{eqnarray}
M_{W_2} & \geq & 477\; \mbox{\rm GeV} \nonumber \\
\xi & \leq & 0.031\; \mbox{\rm rad}
\end{eqnarray}
we obtain (for e($\mu$) transition to $\tau$ neutrino)
\begin{equation}
\mu_{ab} \leq 6 \cdot 10^{-13} \mu_B
\end{equation}
This value is at the edge of physical interest. However, (9)
is very optimistic. For $M_{W_2} >> M_{W_1}$ we have
\begin{equation}
\xi \simeq \epsilon \left( \frac{M_{W_1}}{M_{W_2}} \right)^2
\end{equation}
From the above and (8) it follows that  $\epsilon \simeq 1$
but this value (which is equivalent to $\kappa_1 \simeq \kappa_2$, 
$\kappa_{1,2}$ are VEV for the bidoublet $\phi$) is very
unprobable \cite{gun,eck}. 

\subsection{Diagrams with scalars in the loop}

To calculate the diagrams of Fig.(2) and (3) we need couplings of charged
Higgs particles to gauge bosons and leptons. To this end we must define
the Yukawa and the Higgs sector. The latter is the most general potential 
\cite{gun} with a vanishing VEV for the neutral field of the left-handed
triplet $\Delta_L$, $v_L=0$.  All necessary couplings can be found in
\cite{pr3}.

We found that all couplings vanish in the $\epsilon=\frac{2 \kappa_1 \kappa_2}{\kappa_1^2+\kappa_2^2}
\rightarrow 0$ limit.
For $\epsilon \neq 0$ the magnitude of the Higgs diagram contribution
to $\mu_{\alpha \beta}$ is also very small. For example diagrams of
 Fig.(3)  for $H_2$ and $W_2$
exchange yield
\begin{eqnarray*}
\mu_{ab} & \simeq & \frac{1}{\sqrt{2}\cdot 4\pi^2}f(\epsilon)
\cdot  \left( m^N_a+m^N_b \right) Im(K_L^{\dagger}K_L) \nonumber \\
&\times&  
\frac{m_e}{M_{H_2}^2-M_{W_2}^2}\left( \frac{2M_{H_2}^2}{M_{H_2}^2-
M_{W_2}^2} ln \left(
\frac{M_{H_2}}{M_{W_2}}\right) -1 \right)  \mu_B.    
\end{eqnarray*}
\begin{equation}
\end{equation} 
where
$f(\epsilon)  <1$
and  $f(\epsilon) \rightarrow 0$ for $\epsilon \rightarrow 0$.
Taking  $M_{W_2(H_2)}=1(1.6)$ TeV we have
\begin{equation}
\mu_{ab} \leq f(\epsilon)  \cdot 10^{-22} \left( \frac{ m_a^N+m_b^N}{eV} \right)
\mu_B.
\end{equation}

Let's note that the expectation that the Higgs diagrams contain terms
proportional to the heavy neutrino masses was incorrect.
Only light masses  remain after cancellation due to the Majorana
nature of neutrinos \cite{nasza}.
Taking into account all diagrams (Figs.1-3) we can see that the dominant 
contribution to Majorana neutrino transition magnetic moment $\mu_{\alpha \beta}$ 
is given by Eq.(6).

We also see that $\mu_{\alpha \beta} \neq 0$ if $K_{L(R)}$ matrices are
complex which is the case of broken CP symmetry. When CP is conserved the
magnetic transition is possible only for neutrinos of opposite CP
eigenvalues. For three light neutrinos this leaves two nonzero values at
most.

\section{Conclusion}

We have recalculated the transition magnetic moments of Majorana
neutrinos in the Left-Right symmetric model with a bidoublet and two triplets.
We found that the contribution of the diagrams with Higgs scalar exchange 
is very small,  smaller than the SM contribution (Eq.5).
Only one diagram (with $W_1$ exchange) gives an interesting result  which
for $\mu_{\nu_e  \nu_\mu}$ transition magnetic moment is
\begin{equation}
\mu_{\nu_e  \nu_\mu}/\mu_B \simeq 7 \cdot 10^{-12} |(K_R)_{e \nu_e}|
\left( \frac{0.5 TeV}{M_{W_2}(TeV)} \right)^2 \cdot \epsilon
\end{equation}

For acceptable values of the $K_R$ matrix elements, the mass of the heavy $W_2$
particle and the $\epsilon$ parameter, the value of $\mu_{\nu_e \nu_\mu}$
is however much too small to be interesting from experimental and astrophysical
point of view.

\section*{Acknowledgements}
This work was supported by Polish Committee for Scientific Researches under
Grant No. 2P03B08414 and 2P03B04215.

\end{document}